# RIS Constructing 6G Near-field Networks: Opportunities and Challenges

**Yajun Zhao[1*]**
[1]ZTE Corporation, Beijing 100029, China.
Corresponding author: Yajun Zhao (e-mail: zhao.yajun1@zte.com.cn).

**Abstract:** Near-field propagation, particularly that enabled by reconfigurable intelligent surfaces (RIS), has emerged as a promising research topic in recent years. However, a comprehensive literature review on RIS-based near-field technologies is still lacking. This article aims to fill this gap by providing a brief overview of near-field concepts and a systematic survey of the state-of-the-art RIS-based near-field technologies. The focus is on three key aspects: the construction of ubiquitous near-field wireless propagation environments using RIS, the enabling of new near-field paradigms for 6G networks through RIS, and the challenges faced by RIS-based near-field technologies. This technical review intends to facilitate the development and innovation of RIS-based near-field technologies.

## 1. Introduction

As the commercial scale of 5G wireless networks continues to expand, exploratory research on the next generation of 6G wireless networks is also in full swing. Compared with traditional wireless networks, people have set more ambitious visions for 6G networks and higher performance indicators. In addition, intelligent intrinsic, intelligent wireless environment, integrating sensing and communication, and integration of air, earth, and sea are considered potential key technologies for 6G networks. However, these higher design goals and new technological elements bring many challenges. To address these challenges, research has expanded to the use of high-frequency bands such as millimeter waves and terahertz, further enhancement of multi-antenna technology (e.g., Extremely large-scale antenna array (ELAA), Cell-free), and the introduction of brand-new Reconfigurable Intelligent Surface (RIS)[1][2].

Traditional wireless communication networks (1G~5G) mainly use frequencies below 6GHz, and even below 3GHz. Limited by wavelength, these networks generally use a small number of antenna arrays. Due to the low-dimensional antenna array and low frequency, the wireless near-field range is usually limited to a few meters or even a few centimeters. Therefore, the far-field assumption can be used to design traditional wireless communication systems effectively. However, considering the large aperture and extremely high frequency of ELAA, the 6G network presents a super large near-field area of hundreds of meters, and the traditional far-field plane wave assumption is no longer applicable[3]. Therefore, in the 6G network, the near-field area is not negligible, which stimulates the research of new near-field communication (Near-field Communications, NFC) paradigms.

From the perspective of spatial resource utilization, the typical deployment of traditional cellular networks is a standard network architecture centered on the cell. Under this network architecture, especially in its mainstream sub-6GHz frequency band, the far-field approximation is sufficient as a characterization means. Traditional wireless communication systems have fully exploited and utilized far-field spatial resources. Further exploration and utilization of near-field spatial resources are expected to provide new physical spatial dimensions for wireless communication systems. Future 6G networks will be equipped with larger antenna apertures, and will use higher frequency bands such as millimeter waves and terahertz. This will make the near-field characteristics more significant. At the same time, the introduction of new technologies such as RIS[4][5], ultra-large-scale MIMO, and cell-free (Cell-free)[6] will make near-field scenarios ubiquitous in future wireless networks. Near-field communication technology is also one of the enabling technologies to achieve higher data rate requirements, high-precision sensing demands, and wireless power transmission needs of the IoT in the future 6G network, and has the opportunity to become one of the key technologies of the future 6G wireless air interface. Among them, RIS, with its many characteristics such as large size, passive abnormal adjustment, low cost, low power consumption, and easy deployment, has the opportunity to build a ubiquitous near-field wireless propagation environment in the future 6G network, and bring a new network paradigm.

In recent years, research on near-field propagation characteristics has received much attention, and research progress is rapidly changing. In particular, the field of near-field based on RIS (intelligent metasurface) is a hot research topic, and many research results have emerged. However, there is currently no literature that comprehensively sorts out the near-field technology based on RIS. This article aims to comprehensively and systematically sort out the near-field technology based on RIS from several aspects such as constructing a ubiquitous near-field wireless propagation environment with RIS, enabling new paradigms for 6G networks with RIS, and the challenges faced by near-field technology based on RIS, in order to promote the development of RIS and near-field technology research.

# 2. Near-field Wireless Propagation Concepts

## 2.1 Basic Concepts of Electromagnetic Wave Near-field

According to electromagnetic field and antenna theory, the electromagnetic field radiated by the antenna is divided into near-field region (Near-field region) and far-field region (Far-field region). The near-field region is further divided into reactive near-field region (Reactive near-field region) (also known as non-radiating near-field region, Non-radiating near-field region) and radiating near-field region (Radiating near-field region). In the vicinity of the antenna or scatterer, non-radiating near-field behavior dominates; while in the area away from the antenna or scatterer, electromagnetic radiation behavior dominates. When the antenna radiates signals in free space, the field distribution is uniquely determined by the Maxwell equation, and the propagation characteristics vary in different regions. Electromagnetic waves in these regions show different propagation characteristics. In the far-field region, the changes in amplitude, angle, and phase can be ignored, and the path loss effect is dominant in determining the received signal strength. In the near-field, there are obvious changes in amplitude, angle, and phase according to the distance from the user equipment (UE) to the antenna surface[7][8]. In the far-field region, the plane wave model is used, and the signals on the antenna array are parallel, each antenna has the same arrival angle, and the phase difference between different array elements is only related to the arrival angle. In the near-field region, different antenna signals cannot be regarded as parallel, and the signal arrives at the array in the form of a spherical wave. The phase difference is not only related to the arrival angle but also to the distance. From the perspective of beamforming, beam manipulation includes focusing energy in a specific direction in the far-field (corresponding to focusing at infinity) and allowing energy to be focused on a specific point in space in the near-field [9].

The boundary distance between near-field and far-field is generally called the Rayleigh distance (Rayleigh distance), which is defined as $d = \frac{2D^2}{\lambda}$, where $D$ is the maximum size of the antenna and $\lambda$ is the wavelength [10][11]. The reactive near-field is very close to the antenna surface, and its boundary is considered to be $\lambda/2\pi$, where $\lambda$ is the wavelength. The radiating near-field region (also known as the Fresnel near-field) covers most of the near-field region from $\lambda/2\pi$ to the Rayleigh distance [12].

The motivation for the Rayleigh distance is to consider the phase difference caused by the curvature of the electromagnetic wave front, which is only applicable to the near-axis region near the antenna array and assumes a maximum phase deviation with $\pi/8$ [13]. In fact, the transition between the near-field and far-field is gradual, and there is no strict boundary between the two regions. To define these measures, two different perspectives can be adopted, namely phase error and channel gain error.

1) From the perspective of phase error, some commonly used empirical rules include the Rayleigh distance [13] and the Fraunhofer condition [14], which are mainly applicable to the field boundaries near the main axis of the antenna aperture. Decarli and Dardari et al. extended the near-field and far-field boundaries from $2D^2/\lambda$ to $2D^2\cos^2\vartheta/\lambda$, where $\vartheta$ represents the angle relative to the array axis direction [13] .

2) From the perspective of channel gain error, the field boundary in the off-axis region can be described more accurately. An improved Rayleigh distance, called the effective Rayleigh distance, was proposed in [14]. The authors defined the effective Rayleigh distance based on the standard of normalized coherence greater than 95%, obtaining the effective Rayleigh distance $\varepsilon\cos^2 2D^2\vartheta/\lambda$, $\varepsilon = 0.367$ , where $\vartheta$ is the angle between the center of the ULA-based base station antenna and the UE. The effective Rayleigh distance is smaller than the classical Rayleigh distance and is related to the direction angle $\vartheta$.

Since the reactive near-field region is small (less than the wavelength), the evanescent wave decays rapidly with distance, and the distance between the objects interacting in the 6G system is generally larger than the wavelength. Therefore, for the rest of this article, we will focus on the propagation characteristics in the radiating near-field region, and the term "near-field" will be used to refer to the radiating near-field region without special instructions.

## 2.2 Near-field Propagation Characteristics

Near-field propagation will bring a different channel environment from far-field propagation: near-field effects and spatial non-stationarity [15]. In addition, near-field will also bring a more significant broadband beam squint effect. Compared with traditional far-field, near-field has three significant features, namely spherical wave model (Spherical Wave Model, SWM), spatial non-stationarity (Spatial Non-Stationarity, SNS), and beam squint effect (Beam Squint Effect, BSE). In addition, the near-field fading changes more drastically with distance: the far-field signal strength fades with the square of the distance, while the near-field signal strength fades with the 4th to 6th power of the distance.

**A. Near-field effect - Spherical Wave Model (SWM)**

One of the main features of near-field propagation is the spherical wavefront. In actual wireless communication scenarios, the electromagnetic waves from the antenna array propagate in the form of spherical waves in the radiation field. However, when the distance between the user's receiving antenna and the transmitting antenna is greater than the Rayleigh distance (i.e., in the far-field region), the spherical wave model can be approximated by the plane wave model for convenience in calculations under the predefined minimum deviation constraint. However, in ultra-large-scale antenna array systems, due to the significant increase in the number of antenna elements and the subsequent increase in the array size, the Rayleigh distance at the boundary between the near-field and far-field regions increases, and the near-field region of the array also expands. In this case, the commonly used plane wave model in channel modeling is no longer applicable, and it cannot accurately model the real channel conditions. When the antenna array is large (i.e., significantly larger than the wavelength, such as ELAA and RIS) and operates at short distances, the plane wave approximation is violated, and the SWM (or wavefront curvature) must be considered [18]. Therefore, for ultra-large-scale antenna arrays, the spherical wave model is a more reasonable choice for channel modeling.

In conventional far-field conditions, for the plane wave model used in the far-field region, the signals on the antenna array are parallel, each antenna has the same angle of arrival (AOA), and the phase difference between different antenna elements is only related to the angle of arrival. In the near-field region, different antenna signals cannot be regarded as parallel, and the signals arriving at the array take the form of spherical waves. The phase difference between different antenna elements is not only related to the angle of arrival but also to the distance. Reference [16] proposes the use of the parabolic wave model (Parabolic Wave Model, ParWM) to simulate the spherical wave when modeling the near-field effect. The parabolic wave is a second-order approximation of the spherical wave model, which is simpler than the spherical wave model and avoids the complex root calculations required for spherical wave modeling. From a communication perspective, near-field propagation enhances spatial multiple access capabilities, allowing simultaneous service for users at the same angle but different distances (i.e., beam focusing) [17].

**B. Spatial non-stationarity (Spatial Non-Stationarity, SNS)**

Spatial non-stationarity (Spatial Non-Stationarity, SNS) refers to the variability of wireless channel characteristics in space and is another major difference between near-field and far-field. Spatial non-stationarity can occur in both far-field and near-field. In the far-field, spatial non-stationarity is usually caused by the large aperture of the BS antenna array, where different users may see different parts of the array or encounter different scatterers. In the near-field, in addition to the large array aperture factor, the non-linear phase shift between array elements caused by the spherical wavefront also leads to spatial non-stationarity, making the situation more complex [19]. For the large array aperture factor, due to near-field scattering, local fading, and blockage effects, different antenna elements of the large array observe signals from the same signal source at different angles, and the large array observes the same channel path with different powers or different regions receive different propagation path signals, leading to spatial non-stationarity of the entire array [20][21][22]. That is, SNS includes two cases: (1) different regions of the array may observe the same sub-path signal, but the power falling on different array regions is different; (2) different regions of the array may observe completely different paths. When the array size becomes very large, because each UE's signal energy is concentrated on a part of the array (called the visible region), different parts of the array may also see different UEs. Therefore, due to the limitation of the effective array size by the large array visible region (i.e., the effective array size is limited), the performance of a single UE will not continue to improve with the increase in array size. However, this non-stationary characteristic can greatly enhance the multi-user multiple access capability, which is beneficial for scenarios with a large number of access users, that is, it can better support block multiple access (see Section 4.1.3). Compared with traditional natural scatterers, the beamforming of RIS will make the signal propagation more focused, making the signal propagation vary more dramatically with angle and distance, thus aggravating the spatial non-stationary characteristics.

### C. Broadband Beam Squint Effect (Beam Squint Effect, BSE)

Traditional wireless systems mainly operate in low-frequency narrowband scenarios, so traditional research mainly adopts the narrowband assumption. In the future, high-frequency bands such as millimeter waves will become the main frequency bands for wireless systems, and broadband communication will become the main scenario due to the rich spectrum resources. In broadband communication, the beam squint effect (beam squint effect, BSE) needs to be considered, and near-field propagation will further aggravate the broadband squint effect.

For traditional active phased array antennas, due to the same phase shift applied to different frequency signals by the analog phase shifter, BSE occurs. Similarly, for RIS, the same phase control coefficient is applied to different frequency signals, which also leads to BSE. When the signal bandwidth is large, the signal is incident on the antenna array at an angle, and the optical path difference between the high and low frequency bands is not negligible. The fixed antenna spacing will cause frequency-dependent phase shifts on the array, leading to BSE. In the far-field, BSE causes beam splitting in different directions, while in the near-field scenario, BSE will cause the beams of different frequencies to defocus [23], so that most beams cannot focus on a single target user [24]. For communication, BSE will reduce the beamforming gain, leading to performance degradation. Although the broadband beam squint effect brings challenges, it can also be used to support some specific requirements. For example, using the multiple beam splitting of different subbands due to BSE, multiple UEs can be quickly accessed in different beam splitting, or used for fast beam scanning training. It should be noted that BSE occurs under specific system conditions, that is, large bandwidth, large array size, frequency-independent analog phase shifter array structure, and mainly affects the beamforming/beam focusing angle away from the normal direction.

In dealing with the problem of broadband beam squint, RIS near-field broadband will face new challenges. Traditional active phased array antennas overcome the broadband squint phenomenon by using true time delay (True Time Delay, TTD) mechanism, that is, introducing TTD structure on the active phase shifter. However, RIS is a passive control and it is difficult to configure TTD structure. Therefore, in the design of RIS control matrix, it is necessary to introduce TTD matrix components on the basis of considering the ideal control matrix to overcome the broadband squint effect. Reference [25] further pointed out that TTD matrix components can also be used to achieve independent sub-band control of RIS.

# 3. RIS Constructing Ubiquitous Near-field Wireless Propagation Environment

## 3.1 RIS Constructing Near-field Wireless Propagation Environment

The unique technical features of RIS can be used as an effective means to solve the challenges faced by traditional active phased array antennas. First, RIS, as a programmable two-dimensional electromagnetic metasurface, can abnormally control electromagnetic waves in a passive manner, with the advantages of low cost and low power consumption, and can be easily made into a large antenna aperture, thus realizing dense deployment at a low cost; second, RIS types are diverse and can flexibly adapt to complex and diverse deployment environments. From a functional perspective, RIS types can include channel control type (e.g., reflective RIS, transmissive RIS, and semi-transparent semi-reflective RIS), information modulation type RIS (e.g., RIS-based new base station, RIS-based backscatter transmitter, and RIS-based companion communication), and RIS-based new phased array antennas, etc. RIS can be easily made into different sizes, shapes, and surface forms to meet different deployment requirements; finally, the simple and easy-to-deploy features of RIS can also easily build a near-field LOS environment, thereby better supporting the needs of perception positioning and wireless power transmission. In addition, since RIS is passive control, it naturally has a low level of electromagnetic radiation, which can still meet the human electromagnetic radiation safety index specific absorption rate (Specific Absorption Rate, SAR) in the ubiquitous near-field environment.

In summary, compared with traditional active phased array antennas, RIS has the characteristics of passive control, low cost, and easy deployment, and can be densely and widely deployed, thus having the opportunity to build a ubiquitous near-field channel environment for future 6G networks.

## 3.2 Typical Near-field Modes based on RIS

The introduction of RIS has constructed a cascaded channel, and the wireless propagation environment of future 6G networks will be more complex and diverse compared with traditional networks. From the perspective of near-field propagation environment, the typical near-field modes constructed by RIS can be classified as follows. These scene classifications can be referred to when analyzing the near-field propagation characteristics of RIS.

Table 3.1 Near-field from the perspective of RIS function

| Function type | Near-field Characteristics |
| --- | --- |
| RIS Type for Channel Regulation | ① Extend near-field coverage area<br>② Overcome near-field coverage gaps<br>③ Construct new LOS near-field |
| RIS-based New PAA | Achieve large aperture phased-array antennas with |
| | low cost and complexity, enhancing near-field coverage distance. |
| RIS Type for Information Modulation | For low-rate IoT device communication, build a near-field transmitting environment |

Table 3.2 Near-field from the perspective of transmission and reflection types

| Type | Near-field Characteristics |
|---|---|
| Reflective RIS | Constructing a near-field propagation environment within the positive angle range $(0,\pi)$ of the RIS incident on signal electromagnetic waves |
| Transmission RIS | Constructing a near-field propagation environment within the range of the reverse angle $(\pi,2\pi)$ of the RIS incident on the signal electromagnetic wave |
| Simultaneous Transmitting and Reflecting Reconfigurable Intelligent Surfaces (STAR-RIS) | Construct a near-field propagation environment within the angle range $[0,2\pi)$ |

Table 3.3 Passive/Active RIS near-field

| Type | Near-field Characteristics |
|---|---|
| Passive RIS | Focus the incident electromagnetic wave signal in the near-field, but the signal strength is limited |
| Active RIS | Focus the incident electromagnetic wave signal in the near-field and amplify the signal. It can overcome the problem of limited signal strength in passive RIS, but the complexity is slightly higher |

Table 3.4 Near-field and far-field combinations in RIS networks

| Type | Near-field Characteristics |
|---|---|
| Single RIS cascaded channel (NB-RIS-UE) | NB-RIS channel: near-field/far-field<br>RIS-UE channel: near-field/far-field |
| Multiple RISs cascaded channels (NB-RIS-RIS-UE) | RIS-RIS channel: near-field/far-field<br>NB-RIS channel: near-field/far-field<br>RIS-UE channel: near-field/far-field |
| (RIS cascaded channel) + (NB and UE direct channel) | NB-UE direct channel: near-field/far-field<br>RIS-RIS channel: near-field/far-field<br>NB-RIS channel: near-field/far-field<br>RIS-UE channel: near-field/far-field |

## 3.3 RIS Abnormal Control and Natural Scattering

When RIS is deployed in the wireless propagation channel between the transmitter and receiver, it has the capability to reconstruct the electromagnetic propagation environment, creating an entirely new near-field communication environment. In the natural propagation environment, natural reflection mainly includes specular reflection and natural scattering. RIS introduction will bring three changes to the propagation of electromagnetic waves: abnormal control replacing original specular reflection, abnormal control replacing original natural scattering, and introducing additional propagation paths (and the propagation of this path is abnormally controlled by RIS).

Specular reflection, as a special case of natural scattering, is analyzed separately from general natural scattering. In specular reflection, the distance between the transmitter and receiver is equal to the accumulated distance of segmented reflection paths (i.e., the sum of segmented channel distances). Only when this accumulated distance is less than the Rayleigh distance, the channel propagation characteristics meet the near-field conditions. Obviously, the path component length of specular reflection is longer than the direct LoS path between the transmitter and receiver. In addition, specular reflection does not have beamforming gain.

For general natural scattering, different scattering directions will separate the electromagnetic waves into different angle spaces. In the case of isotropic scattering, the electromagnetic wave signal energy will be evenly scattered into different angles. However, in the case of anisotropic scattering, a larger proportion of energy will be scattered in some angles, while a smaller proportion of electromagnetic wave energy will be scattered in other angles, reflecting the fluctuation of energy distribution in the angle domain. When the UE is in a certain angle region of natural scattering, the gathered scatter rays in this angle can be regarded as the scatterer focusing these scatter rays on the terminal. Although this propagation phenomenon has high spatial freedom, due to the fact that only a small proportion of signal energy is scattered in this angle, the signal strength is weak and the channel capacity is limited. Like specular reflection, general natural scattering also does not have beamforming gain.

RIS introduction realizes abnormal control of signal electromagnetic waves (actively 人为地), thereby adaptively controlling the signal energy to the target UE in the form of beamforming/beam focusing. Unlike specular reflection, RIS has a focusing effect on signal propagation and can change the focus/focal plane of propagation. RIS is equivalent to a zoom lens that can focus on signal propagation and control the signal to different orthogonal subspaces. RIS provides beamforming gain and supports more flexible abnormal control. In addition, in cases where there are no scattering paths in the natural propagation environment, RIS can construct new propagation paths (and the propagation of this path is abnormally controlled by RIS) by properly deploying RIS.

In actual network deployment, RIS can be flexibly deployed in the area close to the UE to construct a near-field propagation environment, which can enhance the signal while fully obtaining the gain of near-field propagation.

# 4. RIS Enables New Paradigms for 6G Network Near-field

RIS possesses several characteristics, including substantial dimensions, passive anomaly regulation, cost-effectiveness, low power consumption, and ease of deployment. In future 6G networks, it holds the potential to establish a pervasive near-field wireless propagation environment, ushering in a novel network paradigm.

## 4.1 Enhanced Wireless Communication - Spatial New Dimension

### 4.1.1 Expand spatial freedom and increase channel capacity

In traditional wireless communication, continuously increasing the number of MIMO's transmit and receive antennas is an effective means to continue the wireless channel capacity. However, this approach will cause channel hardening (Channel hardening) after reaching a certain level, and the rank (Rank) of the channel cannot be further improved. The antenna array elements in the 5G system's Massive MIMO have reached a high level, and channel hardening has occurred, making it difficult to further improve the channel capacity by simply increasing the number of antennas. However, traditional MIMO works at low frequencies and with small antenna apertures, and the near-field region is very small. It is based on the far-field assumption for research, so the channel hardening mentioned is for far-field spatial resources. As mentioned earlier, ubiquitous deployment of RIS can construct a ubiquitous near-field propagation environment, and the near-field propagation characteristics expand new spatial dimensions and increase the freedom of spatial resources.

In the field of wireless communication, freedom degree (Degree of Freedom, DoF) has become a key measure to understand and evaluate system capabilities and potential [25][26]. Simply put, DoF provides a means of the number of independent signal dimensions that can be used to transmit information in the wireless channel. Although traditional far-field scenarios have been extensively studied, near-field shows unique features and requires new exploration of DoF. There are three technical advantages in analyzing near-field from the perspective of DoF: first, NFC provides more DoF, which makes NFC advantageous in terms of data capacity and transmission capability; second, characterizing DoF in NFC helps optimize system parameters such as antenna configuration and transmission strategy, thereby improving overall performance; third, using the DoF perspective helps develop communication protocols and algorithms specifically tailored for the NFC environment, thereby enhancing reliability, coverage, and throughput. Although some studies have analyzed the DoF of NFC, this field is still in its infancy [27].

For single-user access scenarios, the improvement of near-field freedom can overcome the problem of channel rank hardening in traditional far-field Massive MIMO and solve the problem of limited rank under high-frequency LoS paths, thereby improving the near-field spatial multiplexing gain for single-user access. In addition, near-field beam focusing can also improve beamforming gain.

For multi-user access, near-field propagation can enhance the capacity of space-division multiple access (SDMA), thereby improving the system's multi-user access capability. Near-field beam focusing has energy focusing capabilities in the angle-distance domain. That is, near-field SDMA can generate beams with spherical wavefronts and serve users at similar angles but different distances. The distance domain of the spherical wavefront provides a new spatial dimension for multi-user access, and near-field SDMA can also be regarded as location-division multiple access (LDMA) [11]. For example, LDMA supports massive IoT access. In addition, in high-frequency broadband scenarios, the broadband beam splitting/squint effect of near-field is more significant. Fully utilizing the beam splitting phenomenon can be used to support fast beam scanning/training and better support multi-UE access and massive IoT access.

#### 4.1.2 RIS supports continuous coverage of cm/mmWave high-frequency band networks

The frequency bands below 6GHz are already or will be fully occupied by existing 4G/5G commercial networks. Although the ITU has allocated 500MHz of bandwidth in the 6GHz band to IMT [30], and China has officially designated this band for 5G and 6G networks, each operator can only be allocated a narrow bandwidth, which is impossible to meet the larger bandwidth requirements of future 6G. In addition, the millimeter wave band is also a high-quality band for supporting continuous coverage of high-precision sensing and positioning services. It can be expected that the millimeter wave band, especially the centimeter wave and low-frequency millimeter wave frequency bands (i.e., 7-30GHz), will have the opportunity to become the main frequency band for realizing continuous coverage of 6G networks.

The electromagnetic characteristics of the millimeter wave band determine the high directivity of millimeter wave communication. Traditional millimeter wave communication is limited by high path loss and high blockage probability, making it difficult to achieve continuous coverage based on the traditional centralized deployment of base stations (cellular networks). Factors such as site selection, cost, complexity, power consumption, and deployment difficulty will limit the further increase in the deployment density of traditional base stations. The unique technical advantages of RIS, such as low cost, low power consumption, and easy deployment, make it a potential solution for achieving continuous coverage of millimeter wave networks. For example, by simply reusing existing 5G base station sites (i.e., the density of millimeter wave base stations is similar to that of base stations in the 6GHz band or slightly increased) and densely deploying RIS between base stations, continuous coverage of millimeter waves can be achieved at a low cost.

To better build the continuity of near-field coverage in millimeter wave networks, RIS network deployment needs to consider both traditional signal strength coverage and near-field propagation conditions in terms of coverage. Therefore, RIS network deployment needs comprehensive optimization in terms of size, orientation angle, deployment density, and deployment location. That is, RIS network deployment needs to add the spatial freedom of near-field propagation channels as an optimization goal based on the original optimization goal of signal coverage strength. For example, in the target continuous coverage area, the reachable spatial freedom of near-field propagation channels exceeds a predefined threshold (e.g., probability greater than ).

#### 4.1.3 RIS Block Division Multiple Access

When RIS serves multiple users, the beams of different users may simultaneously hit the same RIS panel. Since RIS can only adopt one control state at any given time, it is difficult for RIS beamforming control to optimally match the channel characteristics of multiple users. To better achieve multi-user access for RIS, in addition to using traditional non-orthogonal multiple access (NOMA) technology, a block division multiple access (BDMA) mechanism based on RIS was proposed in literature [5]. The BDMA mechanism divides the RIS into multiple sub-blocks, and the base station (Node B, NB) points the beams of different users to different sub-blocks of the RIS. By independently controlling the phase of each sub-block, it is possible to simultaneously optimize the matching of different users' channels, effectively achieving multi-user access [31].

When the channel between NB and RIS meets the near-field propagation conditions, the beam energy for a specific UE can be better concentrated on a particular sub-block of the RIS, which is more conducive to realizing block division multiple access. In addition, the more significant spatial non-stationarity of near-field propagation can also be used to better support block division multiple access. When the size of the RIS array becomes very large, because the signal energy of each UE is concentrated on a part of the RIS array (referred to as the visible region), different parts of the RIS array may see different UEs. This spatial non-stationarity can greatly enhance the multi-user multiple access capability, which is beneficial for scenarios with a large number of access users, that is, it can better support block division multiple access.

Compared with NOMA technology, BDMA access can effectively eliminate interference between multiple users, thus achieving better performance. Compared with TDMA technology, BDMA access can achieve simultaneous transmission, supporting higher resource scheduling flexibility, thus effectively reducing latency and improving spectrum efficiency. The BDMA access mechanism based on RIS can be applied to both single-cell multi-user access scenarios and multi-cell multi-user access scenarios [32].

## 4.2 Enabling Integrated Sensing and Communications (ISAC)

Integrated Sensing and Communications (ISAC) systems can achieve high-speed communication and high-precision sensing by breaking the traditional separate operation mode of communication and sensing through high-speed and low-latency system information interaction. However, there are differences in the design goals of communication systems and sensing systems. The design goal of the communication system is to transmit information and pursue the ultimate capacity of the wireless channel. The goal of the sensing system is to extract information about the user's state from the wireless channel. As mentioned earlier, 6G networks are expected to use higher frequency bands, wider bandwidths, and ultra-large-scale antenna arrays, which will construct a more ubiquitous near-field propagation environment, helping to improve the performance of wireless communication and sensing systems.

RIS, with its unique technical advantages, provides support for the ubiquitous and high-precision sensing and positioning service requirements. Based on the construction of a near-field propagation environment using RIS, there are several potential benefits for system sensing and positioning:

(1) Utilize the characteristics of the near-field spherical wave model to estimate angle and distance parameters simultaneously. In terms of spatial resolution, near-field beam focusing provides extremely narrow and highly directional beams, which contain both angle and distance information, thus achieving high-resolution parameter estimation. For distance sensing, in traditional far-field propagation conditions, the plane wave model only contains angle information, and the commonly used distance sensing method is to estimate arrival time. However, under near-field propagation conditions, narrowband distance estimation can be achieved using the near-field spherical wave model. The combination of arrival time estimation and distance information estimation contained in the spherical wave propagation can achieve more robust and accurate distance estimation. In addition, compared with the multi-node requirement in the far-field scenario, utilizing the characteristics of the near-field spherical wave model to estimate angle and distance parameters simultaneously can avoid the need for multi-nodes.

(2) Flexibly deployed RIS can easily construct LoS channels to better support sensing and positioning. NLoS (Non-Line of Sight) propagation is considered a major source of error in wireless positioning systems. In traditional NB-centered wireless cellular networks, especially in the millimeter wave band typically used in ISAC systems, LoS paths are easily blocked. Reasonable deployment of RIS can construct new LoS channels, thus supporting precise sensing and positioning even when the direct LoS path between NB and UE is blocked.

(3) Provide higher beamforming gain to enhance the strength of the sensing signal. Compared with traditional sensing systems without RIS, RIS beamforming can enhance the sensing signal and improve the signal-to-noise ratio of sensing, thus enhancing the performance of positioning sensing. For example, RIS can provide propagation paths for NB coverage blind spots and enhance signals in NB weak coverage areas through beamforming, improving the signal-to-noise ratio of sensing signals.

(4) By covering blind spots and constructing ubiquitous near-field environments, improve the coverage rate of sensing and positioning services. As mentioned earlier, the characteristics of RIS technology make it relatively easy to achieve ubiquitous deployment in 6G networks. By optimizing the deployment of RIS to construct a ubiquitous near-field LoS propagation environment, near-field propagation characteristics can be fully utilized for high-precision positioning and sensing, ensuring the coverage of sensing and positioning services in 6G networks.

(5) Overcome multipath interference and improve positioning accuracy. Multipath can be fully utilized to improve the throughput of communication services, but it is an interference factor for positioning sensing [35]. The introduction of RIS can address multipath interference in several ways: (1) The beamforming of large-sized RIS can enhance the cascaded channel, constructing the main propagation path; (2) RIS can be deployed on the main scattering bodies in traditional natural wireless propagation paths; (3) RIS can increase the amount of "geometric" position-related information that can be utilized from the received multipath contour, thereby improving positioning performance.

(6) From a resource scheduling perspective, RIS-based sensing and positioning fully utilize the flexibility of BS signals and the controllability of RIS, which can not only improve the coverage and accuracy of positioning and sensing but also increase the precision.

(7) Utilize near-field effects to support better sensing and positioning. First, utilize the significant broadband beam squint effect in the near-field to support sensing and positioning, mainly by using multiple focusing areas on different subcarriers to accelerate sensing and positioning processes, reduce scanning time, and improve the accuracy and resolution of sensing and positioning; second, utilize the significant spatial non-stationarity in the near-field to support sensing and positioning, that is, estimate channel changes using spatial non-stationarity to better support target detection and positioning.

Of course, ISAC enabled by RIS also faces many challenges, including: (1) the difference in design goals between sensing and positioning and communication; (2) the coordination and scheduling of sensing resources and communication resources; (3) whether the near-far field boundary conditions for communication are consistent with those for sensing and positioning, etc. Literature [31] summarizes the main opportunities, challenges, and possible solutions for positioning and mapping in specific contexts enabled by RIS.

## 4.3 Enabling Simultaneous Wireless Information and Power Transfer (SWIPT)

In future 6G networks, it is expected that a large number of low-power IoT devices will be ubiquitously deployed, and powering these devices will be a huge challenge. Traditional methods such as power lines or removable batteries are no longer suitable, and wireless power transfer (Wireless Power Transfer, WPT) is expected to become a key technology to address this issue [36]. There are three types of technologies to implement WPT: inductive coupling, magnetic resonance coupling, and electromagnetic radiation [37][38]. The first two are more efficient in energy conversion, but they require a very short distance between the power supply and the device, belonging to non-radiative near-field technology. The third type, called radio frequency wireless power transfer (RF-WPT) or remote WPT, is the preferred method for powering remote low-power IoT devices and is the type discussed in this article, focusing specifically on radiative near-field wireless power transfer technology. The radiative near-field region (Fresnel region) is located between the reactive near-field region and the far-field region [39], that is, the distance between the energy transmitter and the energy receiver is greater than the Fresnel limit [40].

Varshney et al. first introduced the concept of Simultaneous Wireless Information and Power Transfer (SWIPT) in 2008 [41]. Due to the dual physical characteristics of electromagnetic waves, SWIPT technology enables the simultaneous transmission of both information and power. In SWIPT systems, the dual functionality of Wireless Power Transfer (WPT) and information transmission is jointly designed on a shared hardware platform, also known as the co-design of communication and energy (CCE). The vision of Tesla and Marconi is envisioned to be integrated into a new unified network paradigm (SWIPT paradigm) in future 6G networks. In this paradigm, radio waves are fully utilized to carry both energy and information. Future network operators may evolve into providers of both telecommunications and energy, enabling them to offer truly ubiquitous wireless connectivity, allowing tens of trillions of future low-power devices to sense, compute, connect, and be powered anywhere and anytime [40].

In the traditional far-field propagation conditions, electromagnetic radiation propagates in the form of plane waves, and the efficiency of WPT is relatively low at this time, with most of the energy being wasted. On the contrary, under radiative near-field propagation conditions, electromagnetic radiation transforms into spherical wave propagation, which enables energy transmission not only to be directed towards the desired direction but also to be focused on specific locations, known as beam focusing. Focused beams can concentrate energy in a particular area, significantly increasing energy density. Beam focusing brings several advantages to radiative near-field WPT systems. Firstly, it improves energy transmission efficiency compared to directional radiation in the far-field. Secondly, the power transmitter can focus the radiated energy on the precise location of the charging device, allowing the receiving device to receive more energy. Furthermore, it can greatly reduce energy pollution and the impact of radiated energy on the human body.

RIS possesses the capability to spatially process energy signals without the need for amplifiers or active phase shifters. Consequently, leveraging RIS enables the efficient transfer of electromagnetic wave energy from one point to another at an exceptionally low cost. By actively controlling the propagation of electromagnetic waves, RIS achieves the efficient transmission of both information and energy. Yu et al. proposed a novel metasurface with multifocal characteristics for efficient Wireless Power Transfer (WPT) [41]. Zhang et al. introduced a new multifocal reflective metasurface for WPT [42].

However, most existing literature primarily focuses on passive RIS-assisted communication systems, which are susceptible to the "double fading" phenomenon in passive RIS-assisted Simultaneous Wireless Information and Power Transfer (SWIPT) systems [46]. To address the impact of the "double fading" effect, active RIS was introduced in references [47][48], which essentially inherits the hardware structure of passive RIS while equipped with a set of low-power amplifiers. Thus, active RIS not only controls the phase of the signal but also amplifies the signal's power. Reference [49] proposed an active RIS-assisted SWIPT system.

Although there is no specific evidence proving adverse health effects, prolonged exposure of the human body to intense radio waves may potentially have detrimental impacts [43]. To prevent such harmful effects, regulatory authorities impose limits on the Specific Absorption Rate (SAR), which measures the human body's absorption of radio waves. The strong electromagnetic field at the beam focus may lead to violations of SAR limits. Therefore, the design and optimization of WPT and WIPT need to consider these safety constraints. As a passive control device, RIS exhibits lower electromagnetic radiation characteristics, ensuring compliance with health and safety SAR indicators while constructing a near-field communication environment.

The design of SWIPT systems requires consideration of several engineering requirements and design challenges: 1) Remote device charging necessitates a broad coverage area; 2) Ensuring a certain end-to-end energy transfer efficiency; 3) Supporting energy transfer in both Line-of-Sight (LoS) and Non-Line-of-Sight (NLoS) scenarios to broaden WPT applications; 4) Supporting a certain level of mobility; 5) Achieving wireless power transfer throughout the network coverage area; 6) Ensuring wireless radiation safety; 7) Reducing the energy consumption of wireless power supply devices; 8) Integrating communication and power transfer (SWIPT); 9) Fusing WPT with 6G sensing, computing, and communication.

# 5. Challenges of RIS Near-field

The dense deployment of RIS constructs a ubiquitous near-field propagation environment, bringing a new network paradigm to future 6G networks. However, this new paradigm also brings many challenges. This section specifically discusses some of the key challenges, such as near-field channel modeling based on RIS, the impact of near-field channels on mobility, and RIS-based network deployment.

## 5.1 Near-field Channel Modeling based on RIS

The introduction of RIS brings new challenges to wireless channel modeling [50][51]. Many teams have conducted research on RIS channel modeling, but most of the published literature is based on far-field assumptions [52][53][54]. For near-field channel models, existing research is mostly focused on active phased array antenna scenarios. The ubiquitous near-field scenario constructed by RIS brings new possibilities for future 6G networks, making it necessary to conduct in-depth research on near-field channel models based on RIS. The study of near-field channel models based on RIS needs to consider the following factors:

(1) Near-field cascaded channels, such as cascaded channels, cascaded and direct mixed channels;
(2) Near-far field mixed channels, such as near-far field mixing of different scatterers, cascaded near-far field mixing;
(3) LOS near-field models, NLOS near-field models, and LOS/NLOS mixed models;
(4) Modeling of spatial non-stationarity in near-field channels;
(5) Modeling of coupling effects between dense RIS antenna arrays;
(6) Quantization modeling of RIS phase/amplitude control;
(7) Modeling of RIS channel reciprocity;
(8) Modeling of different types of RIS, such as reflective, transmissive, semi-transparent, etc.;
(9) Modeling of different RIS shapes, such as rectangular plane/surface, circular plane/surface, etc.;
(10) Multi-RIS scenarios
(11) Multi-RIS cascading, such as multi-hop channels between multiple RIS cascades;
(12) Multi-RIS collaboration, such as synchronization deviation modeling between multiple RIS.
(13) Near-field ISAC channel model based on RIS;
(14) Near-field SWIPT channel model based on RIS.

## 5.2 Near-field Channels and Mobility

Compared with the traditional far-field propagation environment, the RIS-constructed near-field propagation environment more finely divides space and expands spatial freedom in terms of angle and distance, thereby increasing channel capacity, obtaining higher spatial resolution for positioning, and improving the efficiency of wireless energy transfer. However, the fine division of space will also bring huge challenges to mobility.

The near-field propagates from both angle and distance dimensions, and the focal spot region of beam focusing is small. When the UE moves a unit distance, the signal strength will have a larger change rate. This means that, at the same moving speed, the data rate of communication will have a larger change, leading to a deterioration of communication stability. Moreover, since the near-field is divided into space from both angle and distance dimensions, different moving directions of the UE will cause a drastic change in the data rate of communication. Similarly, this phenomenon will also lead to a deterioration of mobile target performance in sensing and energy transfer scenarios.

For the traditional far-field, when moving along the normal direction (distance change), the spatial characteristics remain unchanged, and only the tangential angle change will cause a significant change in spatial characteristics. However, the freedom degree of near-field is strongly related to distance and angle. At different distances from the antenna aperture or at different angles, the freedom degree of near-field will change greatly. When the UE moves or the antenna array rotates, there may be changes in the freedom degree of near-field, or changes between near-field and far-field. For single-user access, the change in channel rank will cause a change in peak data rate; for multi-user access, the drastic change in access capacity may cause the spatial division multiple access (SDMA) capacity of massive IoT services to be insufficient, leading to access failure.

Mobility-induced beam defocusing. The localization process uses ELAA and RIS to focus the antenna/RIS on a small point centered on the user. When the user is moving, it is highly likely that the user will move outside the beam focus in the subsequent time slot. Therefore, the localization process and BS/RIS beamforming configuration must start from scratch (or from the surrounding area of the previous frame's position), resulting in a delay in link re-establishment. For sensing tasks, this mis-focused will result in the loss of tracking the sensing target.

To overcome the mobility issues in near-field propagation, the following aspects can be considered:

(1) Design wider near-field beams to balance the spatial division granularity and mobility. At the same time, wide near-field beams can better support the transmission of broadcast signals;

(2) In the ISAC scenario, measured location information can be used to assist beam dynamic tracking;

(3) Design more efficient and timely CSI measurement and feedback mechanisms to achieve dynamic tracking of RIS beam control;

(4) Design robust codebooks to adapt to dynamic changes;

(5) Design signaling procedures for near-far field cell or regional switching;

(6) Optimize RIS deployment to reduce changes in near-field propagation characteristics, etc.

## 5.3 RIS-based Network Deployment

The dense deployment of RIS creates a ubiquitous near-field propagation environment, bringing a new network paradigm to future 6G networks. However, this new paradigm also presents many challenges. To overcome these challenges, it is necessary to analyze the problem mechanisms in depth and explore the possibility of using AI/large model tools for resource management and scheduling in such a complex RIS near-field network.

**A. Continuous coverage in high-frequency millimeter-wave bands**

As mentioned earlier, millimeter-wave bands are expected to be the core frequency bands for future 6G networks. Millimeter-wave bands can provide sufficient bandwidth, and when combined with ELAA and dense RIS, they can greatly improve network performance. However, the challenges of millimeter-wave deployment cannot be ignored. Due to the propagation characteristics of millimeter waves, they are easily blocked. Although RIS can be used as a key solution, there are still many issues that need to be analyzed and studied regarding simple and low-cost deployment.

**B. Changes in deployment optimization objectives**

For traditional far-field assumption-based cellular network deployment, the main optimization objective is the signal strength distribution of network coverage, also known as link budget. However, for near-field assumption-based network deployment, the optimization objective will not only include the signal strength distribution of coverage but also consider the changes in spatial freedom caused by near-field propagation characteristics. Therefore, when optimizing the deployment of near-field networks, factors such as near-field distance conditions, RIS size, deployment density, and angles relative to the coverage area must be considered.

**C. Challenges of integrated communication-sensing-energizing networks**

As mentioned earlier, future 6G networks will be integrated communication-sensing-energizing networks, greatly expanding the capabilities of wireless networks. However, the three types of services - communication, sensing, and energizing - have significant differences, and these differences will bring huge challenges to network deployment. Firstly, the distribution of services is different. The communication service targets various communication terminals connected to the network; the sensing targets include not only some communication terminals but also many non-communication targets; and the energizing targets are mainly low-power IoT devices. The geographical distribution of the three types of services is quite different, and their service characteristics are also very different. Secondly, the ideal propagation channel for communication services is NLOS, while the ideal propagation channels for sensing and energizing services are LOS channels. In addition, the integrated waveform design for communication, sensing, and energy transfer is challenging, as there are differences in the waveform requirements for information transmission, sensing positioning, and wireless energy transfer. How to meet the differentiated needs of the three types of services and achieve low-cost and low-complexity RIS near-field network deployment will face significant challenges.

### D. Challenges of RIS ubiquitous deployment

RIS is ubiquitously deployed in complex and diverse actual environments and requires a larger size to build a ubiquitous near-field propagation environment. The RIS form needs to be flexible and diverse to adapt to complex and diverse deployment conditions, which creates a contradiction between the diverse deployment environment requirements and the unified standard engineering requirements. Incident angle differences, equivalent antenna aperture changes, and other factors will affect the near-field coverage area, which will impact the RIS deployment location/angle, optimal shape, deployment density, etc. In addition, the ubiquitous deployment of RIS will also bring many challenges to power supply, synchronization, and network control. The diversity of RIS types provides the possibility for solving ubiquitous deployment. Channel regulation types include reflective RIS, transmissive RIS, and STAR-RIS, which can flexibly increase LOS channels and expand near-field coverage areas; RIS-based enhanced phased array antennas can expand antenna apertures at low cost and low complexity, better meeting the near-field propagation conditions at the transmitter and receiver; and RIS-based information regulation makes it more convenient to achieve near-field multi-address access between RIS and IoT terminals.

## 6. Conclusion

This paper discusses the potential of RIS in improving the performance of 6G wireless networks, especially in the near-field area. In 6G networks, near-field communication will become crucial, and the traditional plane wave-based MIMO radiation model will no longer be suitable for near-field communication, bringing new challenges and opportunities for 6G wireless networks. This paper first briefly introduces the basic concepts of near-field and then systematically reviews the progress and challenges of RIS-based near-field technology from three aspects: constructing a ubiquitous near-field wireless propagation environment, enabling new paradigms for 6G network near-field, and the challenges faced by RIS-based near-field technology. RIS has broad research prospects in building 6G near-field networks and deserves more attention and exploration. It may bring significant benefits to 6G wireless networks in terms of performance improvement, cost reduction, and application expansion. Through the technical review in this paper, we hope to promote the development of RIS and near-field technology research.

## References

[1] ITU-R WP 5D, "Framework and overall objectives of the future development of IMT for 2030 and beyond", Sept. 26, 2023.
[2] IMT-2030 (6G) Promotion Group, "6G Typical Scenarios and Key Capabilities" White Paper, July 2022.
[3] Z. Zhou, X. Gao, J. Fang, and Z. Chen, "Spherical wave channel and analysis for large linear array in LoS conditions," in Proc. IEEE Globecom Workshops2015, pp. 1 - 6.
[4] Tiejun Cui, Shi Jin, Jiayi Zhang, Yajun Zhao, Yifei Yuan, Huan Sun, et al. "Research Report on Reconfigurable Intelligent Surface Technology" [R], IMT-2030 (6G) Promotion Group, 2021.
[5] RIS TECH Alliance, Reconfigurable Intelligent Surface White Paper (2023), March 2023, Hangzhou, China, (doi: 10.12142/RISTA.202302002). Available: http://www.risalliance.com/en/riswp2023.html.
[6] Jiayi Zhang, Jiying Xiang, Bo Ai, Mengnan Jian, Yajun Zhao. "6G Multiple Antennas and Intelligent Metasurfaces," Publishing House of Electronics Industry, 2023.
[7] S. Hu, M. C. Ilter and H. Wang, "Near-Field Beamforming for Large Intelligent Surfaces," 2022 IEEE 33rd Annual International Symposium on Personal, Indoor and Mobile Radio Communications (PIMRC), Kyoto, Japan, 2022, pp. 1367-1373, doi: 10.1109/PIMRC54779.2022.9977582.
[8] E. Björnson, T. Ozlem and L. Sanguinetti, "A primer on near-field beamforming for arrays and reconfigurable intelligent surfaces", preprint, 2021.
[9] P. Nepa and A. Buf, ``Near-eld-focused microwave antennas: Near-field shaping and implementation," IEEE Antennas Propag. Mag., vol. 59, no. 3, pp. 42-53, Apr. 2017.
[10] M. Cui and L. Dai, "Channel Estimation for Extremely Large-Scale MIMO: Far-Field or Near-Field?," IEEE Trans. Commun., vol. 70, no. 4, Jan. 2022, pp. 2663–77.
[11] M. Cui, Z. Wu, Y. Lu, X. Wei and L. Dai, "Near-Field MIMO Communications for 6G: Fundamentals, Challenges, Potentials, and Future Directions," in IEEE Communications Magazine, vol. 61, no. 1, pp. 40-46, January 2023, doi: 10.1109/MCOM.004.2200136.
[12] K. T. Selvan and R. Janaswamy, "Fraunhofer and Fresnel Distances: Unified Derivation for Aperture Antennas," IEEE Antennas Propag. Mag., vol. 59, no. 4, Aug. 2017, pp. 12–15.
[13] C. A. Balanis, Antenna Theory: Analysis and Design. John wiley & sons, 2015.
[14] M. Cui, L. Dai, R. Schober, et al. Near-field wideband beamforming for extremely large antenna array. arXiv preprint arXiv:2109.10054, 2021.
[15] A. Lapidoth, S. Moser. Capacity bounds via duality with applications to multiple-antenna systems on flat-fading channels. IEEE Transactions on Information Theory, 2003, 49(10):2426–2467.
[16] X. Cheng, K. Xu, J. Sun and S. Li, "Adaptive Grouping Sparse Bayesian Learning for Channel Estimation in Non-Stationary Uplink Massive MIMO Systems," in IEEE Transactions on Wireless Communications, vol. 18, no. 8, pp. 4184-4198, Aug. 2019, doi: 10.1109/TWC.2019.2922913.

[17] Grant A. Joint decoding and channel estimation for linear MIMO channels[C]. Wireless Communications & Networking Confernce, IEEE, 2000: 1009-1012.
[18] H. Zhang, N. Shlezinger, F. Guidi, D. Dardari, and Y. C. Eldar, “6G wireless communications: From far-field beam steering to near-field beam focusing,” IEEE Commun. Mag., 2023.
[19] D. Dardari, N. Decarli, A. Guerra, and F. Guidi, “LOS/NLOS near-field localization with a large reconfigurable intelligent surface,” IEEE Trans. Wireless Commun., vol. 21, no. 6, pp. 4282–4294, Jun. 2022.
[20] Sun S, Li R, Liu X, et al. How to Differentiate between Near Field and Far Field: Revisiting the Rayleigh Distance[J]. arXiv preprint arXiv:2309.13238, 2023.
[21] J. Tian, Y. Han, S. Jin, and M. Matthaiou, “Low-overhead localization and VR identification for subarray-based ELAA systems,” IEEE Wireless Commun. Lett., vol. 12, no. 5, pp. 784–788, May. 2023.
[22] E. De Carvalho, A. Ali, A. Amiri, M. Angjelichinoski, and R. W. Heath, “Non-stationarities in extra-large-scale massive MIMO,” IEEE Wireless Commun., vol. 27, no. 4, pp. 74–80, Aug. 2020.
[23] Gao X, Tufvesson F, Edfors O, et al. Measured propagation characteristics for very-large MIMO at 2.6GHz[C]// Signals, Systems and Computers (ASILOMAR), 2012 Conference Record of the Forty Sixth Asilomar Conference on. IEEE, 2012.
[24] M. Cui, Z. Wu, Y. Lu, X. Wei, and L. Dai, “Near-field communications for 6G: Fundamentals, challenges, potentials, and future directions,” IEEE Commun. Mag., vol. 61, no. 1, pp. 40–46, Sep. 2022.
[25] The Far-/Near-Field Beam Squint and Solutions for THz Intelligent Reflecting Surface Communications
[26] Yajun ZHAO, Reconfigurable intelligent surfaces for 6G: applications, challenges, and solutions. Frontiers of Information Technology & Electronic Engineering, 24(12), 2023. https://doi.org/10.1631/FITEE.2200666
[27] Ouyang C, Liu Y, Zhang X, et al. Near-field communications: A degree-of-freedom perspective[J]. arXiv preprint arXiv:2308.00362, 2023.
[28] D. Tse and P. Viswanath, Fundamentals of Wireless Communication. Cambridge, U.K.: Cambridge Univ. Press, 2005.
[29] Y. Liu et al., “Near-field communications: A tutorial review,” arXiv preprint arXiv:2305.17751, Accessed on 25 Jul. 2023.
[30] GSMA, The 6 GHz Ecosystem: Demand Drives Scale, August 2022. https://www.gsma.com/spectrum/wp-content/uploads/2022/08/6-GHz-IMT-Ecosystem.pdf
[31] Y. Zhao and X. Lv, "Network Coexistence Analysis of RIS-Assisted Wireless Communications," in IEEE Access, vol. 10, pp. 63442-63454, 2022, doi: 10.1109/ACCESS.2022.3183139.
[32] M. Jian and Y. Zhao, "A Modified Off-grid SBL Channel Estimation and Transmission Strategy for RIS-Assisted Wireless Communication Systems," 2020 International Wireless Communications and Mobile Computing (IWCMC), Limassol, Cyprus, 2020, pp. 1848-1853, doi: 10.1109/IWCMC48107.2020.9148537.
[33] Chen H, Keskin M F, Sakhnini A, et al. 6G Localization and Sensing in the Near Field: Fundamentals, Opportunities, and Challenges[J]. arXiv preprint arXiv:2308.15799, 2023.
[34] H. Wymeersch, J. He, B. Denis, A. Clemente, and M. Juntti, “Radio localization and

mapping with reconfigurable intelligent surfaces: Challenges, opportunities, and research directions," IEEE Vehicular Technology Magazine, vol. 15, no. 4, pp. 52–61, 2020.
[35] O. Rinchi, A. Elzanaty and M. -S. Alouini, "Compressive Near-Field Localization for Multipath RIS-Aided Environments," in IEEE Communications Letters, vol. 26, no. 6, pp. 1268-1272, June 2022, doi: 10.1109/LCOMM.2022.3151036.
[36] N. M. Tran, M. M. Amri, J. H. Park, D. I. Kim and K. W. Choi, "Reconfigurable-Intelligent-Surface-Aided Wireless Power Transfer Systems: Analysis and Implementation," in IEEE Internet of Things Journal, vol. 9, no. 21, pp. 21338-21356, 1 Nov.1, 2022, doi: 10.1109/JIOT.2022.3179691.
[37] Y. Zeng, B. Clerckx, and R. Zhang, "Communications and signals design for wireless power transmission," IEEE Trans. Commun., vol. 65, no. 5, pp. 2264–2290, 2017.
[38] A. Costanzo and D. Masotti, "Energizing 5G: Near- and far-field wireless energy and data transfer as an enabling technology for the 5G IoT," IEEE Microw. Mag., vol. 18, no. 3, pp. 125–136, 2017.
[39] D. R. Smith, V. R. Gowda, O. Yurduseven, S. Larouche, G. Lipworth, Y. Urzhumov, and M. S. Reynolds, "An analysis of beamed wireless power transfer in the fresnel zone using a dynamic, metasurface aperture," Journal of Applied Physics, vol. 121, no. 1, p. 014901, 2017.
[40] C. A. Balanis, Antenna Theory: analysis and design. New Jersey, USA: Wiley, 2016.
[41] L. R. Varshney, "Transporting information and energy simultaneously," in 2008 IEEE Int. Symposium on Information Theory, 2008, pp. 1612–1616.
[42] X. Lu, P. Wang, D. Niyato, D. I. Kim and Z. Han, "Wireless networks with RF energy harvesting: A contemporary survey", IEEE Commun. Surveys Tuts., vol. 17, no. 2, pp. 757-789, Second quarter 2015.
[43] B. Clerckx, J. Kim, K. W. Choi and D. I. Kim, "Foundations of Wireless Information and Power Transfer: Theory, Prototypes, and Experiments," in Proceedings of the IEEE, vol. 110, no. 1, pp. 8-30, Jan. 2022, doi: 10.1109/JPROC.2021.3132369.
[44] Y. Gu and S. Aïssa, "Interference aided energy harvesting in decode-and-forward relaying systems", Proc. IEEE Int. Conf. Commun. (ICC), pp. 5378-5382, 2014.
[45] C. Masouros, "Harvesting signal power from constructive interference in multiuser downlinks" in Wireless Information and Power Transfer: A New Paradigm for Green Communications, New York, NY, USA:Springer, pp. 87-122, 2018.
[46] K. Zhi, C. Pan, H. Ren, K. K. Chai and M. Elkashlan, "Active RIS versus passive RIS: Which is superior with the same power budget?", IEEE Commun. Lett., vol. 26, no. 5, pp. 1150-1154, May 2022.
[47] Z. Zhang et al., "Active RIS vs. passive RIS: Which will prevail in 6G?", arXiv:2103.15154, 2021.
[48] R. Long, Y.-C. Liang, Y. Pei and E. G. Larsson, "Active reconfigurable intelligent surface-aided wireless communications", IEEE Trans. Wireless Commun., vol. 20, no. 8, pp. 4962-4975, Aug. 2021.
[49] H. Ren, Z. Chen, G. Hu, Z. Peng, C. Pan and J. Wang, "Transmission Design for Active RIS-Aided Simultaneous Wireless Information and Power Transfer," in IEEE Wireless Communications Letters, vol. 12, no. 4, pp. 600-604, April 2023, doi: 10.1109/LWC.2023.3235330.
[50] ZHAO Y J, JIAN M N, Applications and challenges of Reconfigurable Intelligent Surface for

6G networks (In Chinese), Radio Communications Technology, 1-16 [2021-08-10]. http://kns.cnki.net/kcms/detail/13.1099.TN.20210805.1107.002.html.
[51] Tiejun Cui, Shi Jin, Jiayi Zhang, Yajun Zhao, Yifei Yuan, Huan Sun, et al. "Research Report on Reconfigurable Intelligent Surface Technology" [R], IMT-2030 (6G) Promotion Group, 2021.
[52] B. Xiong et al., "A Statistical MIMO Channel Model for Reconfigurable Intelligent Surface Assisted Wireless Communications," in IEEE Transactions on Communications, vol. 70, no. 2, pp. 1360-1375, Feb. 2022, doi: 10.1109/TCOMM.2021.3129926.
[53] W. Tang et al., "Wireless communications with programmable metasurface: New paradigms opportunities and challenges on transceiver design", IEEE Wireless Commun., vol. 27, no. 2, pp. 180-187, Apr. 2020.
[54] E. Basar, I. Yildirim and F. Kilinc, "Indoor and Outdoor Physical Channel Modeling and Efficient Positioning for Reconfigurable Intelligent Surfaces in mmWave Bands," in IEEE Transactions on Communications, vol. 69, no. 12, pp. 8600-8611, Dec. 2021.

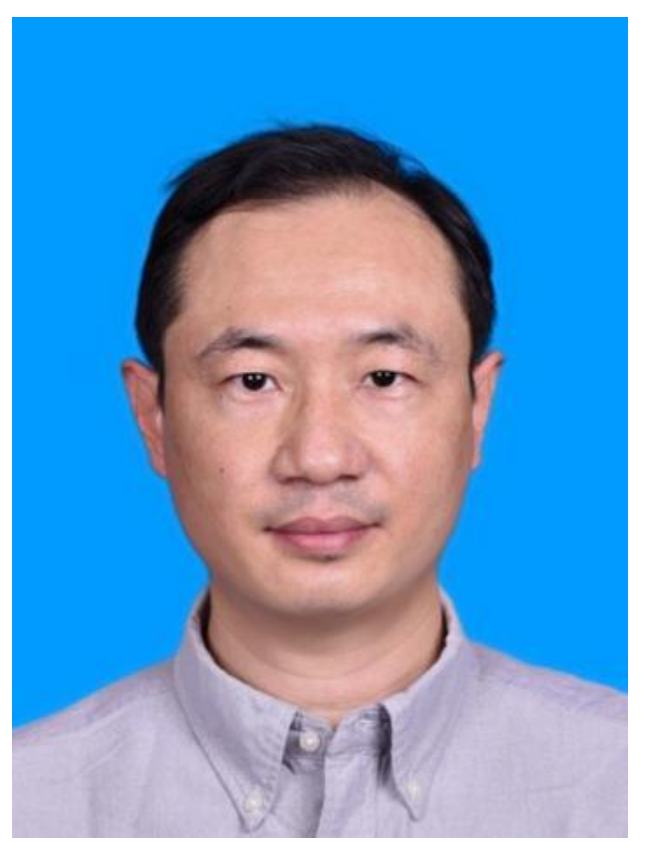

**Zhao Yajun** received the BE, MS, and PhD degrees. Since 2010, he has assumed the role of Chief Engineer within the Wireless and Computing Product R&D Institute at ZTE Corp. Prior to this, he contributed to wireless technology research within the Wireless Research Department at Huawei. Currently, his primary focus centers on 5G standardization technology and the advancement of future mobile communication technology, particularly 6G. His research pursuits encompass a broad spectrum, including reconfigurable intelligent surfaces (RIS), spectrum sharing, flexible duplex, CoMP, and interference mitigation. He has played an instrumental role in founding the RIS Tech Alliance (RISTA) and currently holds the position of Deputy Secretary General within the organization. Additionally, he is a founding member of the RIS task group under the purview of the China IMT-2030 (6G) Promotion Group, where he serves as the Deputy Leader. Email: zhao.yajun1@zte.com.cn